\documentclass[journal]{IEEEtran}
\usepackage{xcolor}
\usepackage{amsmath}
\usepackage{amssymb}
\usepackage{multirow}
\usepackage{booktabs}
\usepackage{colortbl}
\usepackage{graphicx}
\usepackage{tikz}
\usepackage{rotating}

\begin{document}

\title{Forced Oscillations in Power Systems Induced by Data Centers Hosting AI Workloads}

\author{
    Jovan~Krajacic,~\IEEEmembership{Student Member,~IEEE,}
    Georgia~Pierrou,~\IEEEmembership{Member,~IEEE,}
    Maitraya~Desai,~\IEEEmembership{Student Member,~IEEE,} \\
    Gabriela~Hug,~\IEEEmembership{Senior~Member,~IEEE,}
    and~Gustavo~Valverde,~\IEEEmembership{Senior~Member,~IEEE} 
    \thanks{J. Krajacic, M. Desai, G. Hug, and G. Valverde are with the Power Systems Laboratory, ETH Zurich, Switzerland. Emails: \{jkrajacic, mdesai, ghug, gustavov\}@ethz.ch. G. Pierrou is with the Department of Electrical and Computer Engineering at the University of Toronto, Canada. Email: georgia.pierrou@utoronto.ca }
}

\maketitle

\begin{abstract}
Power swings in large Data Centers (DTCs) running Artificial Intelligence (AI) workloads can excite poorly damped modes in power systems. The resulting forced oscillations can lead to flicker, equipment disconnection, or blackouts. This paper investigates the risks of such load fluctuations for different system strengths and damping conditions. Using an analytical approach based on transfer functions, we identify critical DTC locations in the power grid at which load fluctuations could induce the largest forced oscillations and further characterize the harmonic spectrum of the resulting system response. Depending on the frequency and magnitude of the DTC load fluctuations, forced oscillations can become unbounded. The underlying instabilities are classified into saddle-node bifurcations of the forced periodic response and impasse-surface encounters, using Floquet multipliers and the minimum singular value of the algebraic Jacobian. Furthermore, the impact of different duty cycles and harmonic components beyond the fundamental oscillation frequency in square-wave load profiles is analyzed. Finally, the interaction of two oscillating DTCs is investigated for different locations and forcing frequencies, considering both synchronized and unsynchronized operation. The findings can help system operators to define new regulations on the maximum load fluctuations permitted for DTC facilities at specific grid locations, without negatively affecting the stability and operation of the system.
\end{abstract}


\IEEEpeerreviewmaketitle

\section{Introduction}

The advances in Artificial Intelligence (AI) technologies and the growing demand for their services are prompting leading AI companies to build and operate large Data Centers (DTCs). According to the International Energy Agency, over 11,000 DTCs were registered worldwide as of early 2024~\cite{wea24}. The electricity consumption from these loads in 2022 was estimated to be around 1\% of the total electricity consumption (excluding data networks and crypto mining). In the USA, server energy usage increased from about 30~TWh in 2014 to nearly 100~TWh in 2023 due to the proliferation of AI~servers~\cite{Shehabi24}.

The typical size of a modern DTC used for AI model training is in the order of several tens to a few hundred megawatts, and DTCs are expected to reach one or two gigawatts in the near future. Moreover, due to the nature of AI workloads, the load in DTC facilities may fluctuate from operational to almost idle power, back and forth, resulting in periodic demand changes~\cite{Chala25}. In the power system literature, they are known as cyclic loads~\cite{Vanness66,Rao88}.

Depending on the frequency of the oscillations of large cyclic loads, they can excite poorly damped natural modes in the power system, resulting in large sustained oscillations~\cite{Rosta94}. In this regard, the North American Electric Reliability Corporation (NERC) recently expressed concerns about the potential onset of forced oscillations due to AI workloads~\cite{NERC25}. Decades ago, the problem was experienced with other types of large pulsating loads, for example, nuclear accelerators~\cite{Rosta94, Pinneilo71}, arc furnaces, and smelters~\cite{Smolleck91}. Now, the rapid expansion of DTCs may further exacerbate this issue.  

Since the behavior of a forced oscillation is determined by the external input and the internal system features~\cite{Chen23}, the severity of the oscillations depends on system characteristics, such as damping, DTC load characteristics (the magnitude and frequency of the power fluctuations), and the DTC location. The latter is significantly influenced by the proximity to population centers, business ecosystems, network infrastructure, electricity costs, and financial incentives~\cite{wea24},~\cite{Shehabi24}. For instance, in the USA, fifteen states account for 80\% of the national DTC load demand~\cite{EPRI24}. In Europe, DTCs in the Nordic countries benefit from lower electricity costs, while Ireland, with low corporate tax rates, has favored the construction and operation of DTCs~\cite{Kez20,iea24}. Therefore, it is reasonable to expect multiple DTCs in the same grid, resulting in power swings in the order of tens or hundreds of megawatts~\cite{Choukse25}.

In contrast to malfunctioning devices that may cause forced oscillations~\cite{Chen23,Ye17}, DTCs cannot be disconnected. Hence, it is crucial to evaluate the risks associated with AI workloads and the onset of large sustained oscillations. Such conditions may reduce power transfer capability, lead to equipment disconnection, or even blackouts~\cite{YChen25}.

Recent studies on forced oscillations focus on identifying the source location~\cite{YChen25}, including the complex dissipating energy flow method~\cite{Estevez22,Masle24} and a purely data-driven approach in~\cite{Cai25}. However, the location of large DTCs running AI workloads, as potential sources of forced oscillations, is known. Therefore, our interest is to characterize the harmonic spectrum of oscillations analytically for a given load fluctuation, investigate the impact of different load profiles, and identify critical DTC locations in the power grid. System operators may leverage this information to define limits on the maximum load fluctuations in DTC facilities or define stricter rules, such as ramp specifications on critical grid locations~\cite{Choukse25}.

Previous work on evaluating forced oscillations induced by AI workloads is scarce. Among the few works, the authors in~\cite{Ko25} examined the amplitude and variability of forced oscillations considering different DTC penetration levels and sizes, geographical distributions, and changes in system inertia. However, the work did not assess different load patterns, and did not identify critical grid locations where the DTCs could potentially have the highest impact on the grid. In~\cite{Biswas25}, a simulation-based approach is proposed to identify critical grid locations by placing the DTCs in predefined buses and checking the oscillation amplitudes. In large-scale systems, this approach becomes computationally demanding.

In this paper, an analytical framework to evaluate potential onsets of forced oscillations induced by AI workloads is proposed. A comprehensive analysis is conducted and verified by time-domain simulations of DTCs in three test systems. The contributions of the work are:
\begin{itemize}
    \item Impact evaluation under different system and DTC conditions, such as system strength, as well as duty cycles and harmonic components beyond the fundamental oscillation frequency in square-wave load profiles of DTC facilities.
    \item Identification of DTC load fluctuation magnitudes that could result in unbounded oscillations in an otherwise small-signal stable system.
    \item Classification of instabilities into saddle-node bifurcations of the forced periodic response and impasse-surface encounters using Floquet multipliers and the minimum singular value of the algebraic Jacobian, respectively.
    \item Linear analytical derivation of harmonic-component amplitudes in system quantities under DTC-induced forced oscillations, validated against time-domain simulations. 
    \item Identification of critical DTC locations in the power grid.
    \item Impact assessment of multiple DTCs with synchronized and unsynchronized AI workload oscillations.
\end{itemize}

The remainder of the paper is organized as follows: Section II characterizes the load fluctuations observed in DTC facilities during large-scale AI model training and describes the DTC model used in the stability studies; Section III analytically analyzes the impact of these fluctuations on the power system. Finally, Section IV investigates the impact of DTCs in three different test systems, and Section V provides the conclusions.

\section{Data Center Load Profiles and Modeling}

\subsection{Load Profile Analysis}
The servers of a DTC facility, which account for 60\% to 90\% of the total electricity consumption of the DTC, are equipped with Central Processing Units (CPUs) and accelerators, such as Graphics Processing Units (GPUs). These enable AI model training and inference to be split into subproblems for greater efficiency and scalability.
During large model training, thousands of GPUs simultaneously increase and decrease their power demand due to different training phases (forward/backward propagation and synchronization)~\cite{LiLi25}, or because all GPUs are waiting for checkpoints or collective communications to finish~\cite{Llama24}. This may result in sustained square-wave demand fluctuation on the order of tens of MW with frequencies ranging from 0.05 to 1~Hz~\cite{Choukse25}. Indeed, sustained square-wave load profiles have typically been utilized in the absence of real-world data to account for power demand fluctuations from DTCs~\cite{Milano26},~\cite{Ryan25}.

The duty cycle of the AI workload is defined as:
\begin{equation}
    D = \frac{T_{\text{on}}}{T_{f}},
\end{equation}
where $T_{\text{on}}$ is the active duration within one cycle and $T_{f}$ is the total cycle duration, both in seconds. AI-training load cycles may also be bi-periodic~\cite{Biswas25}, where fast fluctuations in the range of 5 to 30~Hz occur in the $T_{\text{on}}$ phase of the low-frequency fluctuation. In model inference, the workload may be distributed among a smaller number of synchronized servers, leading to smaller load fluctuations.

\begin{figure}[t!]
    \centering
    \includegraphics[width=0.95\columnwidth]{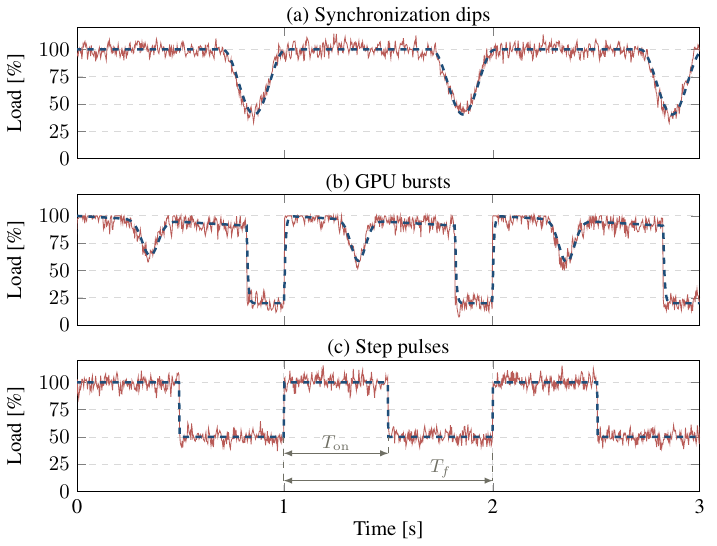}
    \vspace{-0.4cm}
    \caption{Representative DTC load profiles: (a)~synchronization dips, (b)~GPU burst cycles, and (c)~step-pulse square wave. Each profile shows the real-world trace (red, including high-frequency noise) and the filtered trace (dashed blue).
    }
    \vspace{-0.4cm}
    \label{fig:data_center_demand}
\end{figure}

In this paper, we focus on the low-frequency demand fluctuations from AI model training that can excite electromechanical modes. Fig.~\ref{fig:data_center_demand} illustrates representative DTC load profiles drawn from three characteristic AI-training scenarios: (a)~synchronization dips, arising from periodic collective-communication phases during which all GPUs reduce their power consumption while waiting for data synchronization; (b)~GPU burst cycles, characterized by alternating high-power compute phases and short low-demand idle intervals; and (c)~a step-pulse profile, representative of the sustained square-wave demand fluctuation. Each DTC load profile in Fig.~\ref{fig:data_center_demand} is shown in two forms: the real-world trace in solid red, which includes high-frequency load fluctuations, and the filtered trace in dashed blue, which retains the dominant low-frequency variations relevant for this study, and suitable for phasor-domain (RMS) power-system simulations~\cite{RMS_EMT}.

\subsection{Load Profile Modeling}
The filtered step-pulse profile in Fig.~\ref{fig:data_center_demand}(c) is modeled as a square-wave load variation with an adjustable duty cycle, following the approach in~\cite{Milano26, Ryan25}. Specifically, the profile in Fig.~\ref{fig:data_center_demand}(c) varies between 50\% and 100\% of the rated load at a frequency of 1.0~Hz. Even larger demand variations, ranging from 10\% to 100\% of the rated load, have been reported in~\cite{Choukse25}. Since server racks in data centers typically use power-factor-correction devices~\cite{Sun22}, we assume unity power factor and model the IT load demand as active power variation only. Accordingly, the filtered step-pulse profile $P_{\mathrm{load}}$ in Fig.~\ref{fig:data_center_demand}(c) can be represented as a square waveform with duty cycle $D$ and period \(T_f=1/f_f\). Assuming that the load equals \(P_0+P_f\) during the interval \(0\leq t<DT_f\) and \(P_0-P_f\) during the remainder of each period, its Fourier series is:
\begin{equation}
    P_{\mathrm{load}}(t) = P_0 - P_f(1 - 2D) + \sum_{h=1}^{\infty} A_h \cos\!
    \left(h\omega_f t + \theta_h\right),
    \label{eq:fourier_series}
\end{equation}
where $P_f$ is half the peak-to-peak swing of the IT load, $\omega_f=2\pi f_f$ is the angular forcing frequency, while the term $-P_f(1-2D)$ captures the shift in the mean load level relative to $P_0$. For a symmetric waveform, $D = 0.5$ and the mean DTC consumption equals $P_0$. Moreover, the amplitude and phase of each harmonic $h$ are given as~\cite{Oppenheim1997}:
\begin{equation}
    A_h = \frac{4P_f}{h\pi}\bigl|\sin(h\pi D)\bigr|,
    \label{eq:harmonic_amplitude}
\end{equation}
\begin{equation}
    \theta_h = \begin{cases}
        -h\pi D,       & \text{if } \sin(h\pi D) \geq 0, \\
        -h\pi D + \pi, & \text{otherwise.}
    \end{cases}
    \label{eq:harmonic_phase}
\end{equation}

\subsection{Data Center Model and Grid Interconnection}
\begin{figure}[b!]
    \centering
    \vspace{-0.4cm}
    \includegraphics[width=0.95\columnwidth]{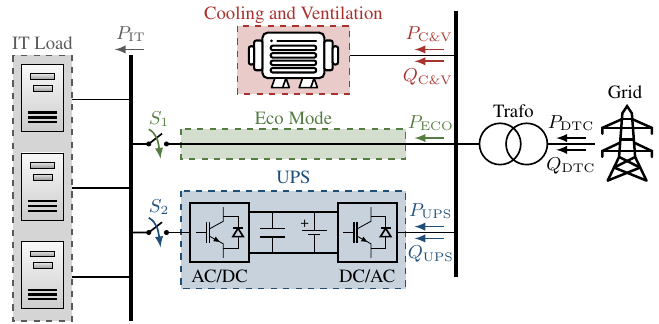}
    \vspace{-0.2cm}
    \caption{Simplified DTC grid-interconnection architecture considered in this work. In Eco mode, switch $S_1$ is closed and the IT load is supplied directly from the grid, bypassing the UPS. In UPS mode, switch $S_2$ is closed and the IT load is supplied through the AC/DC--DC/AC conversion stage. The cooling load remains directly connected to the grid in both operating modes.}
    \label{fig:ups}
\end{figure}

The DTC model considered in this work is shown in Fig.~\ref{fig:ups}. It includes an Uninterruptible Power Supply (UPS), an induction motor to represent the cooling load, and a pulsing IT load that represents the transients caused by AI workloads~\cite{Milano26}. The IT load can be supplied either directly from the grid or through the UPS. In Eco mode, switch $S_1$ is closed and the UPS is bypassed, such that the IT load is connected directly to the grid. Conversely, when $S_1$ is open and $S_2$ is closed, the IT load is supplied through the UPS. The cooling load is connected directly to the grid in both cases, unless otherwise~stated.

The UPS is represented by two average-value converter models connected through a common DC link. The load-side converter supplies the IT load, whereas the grid-side converter controls the power exchanged with the grid. The detailed dynamics of the load-side converter are neglected and approximated by a first-order response. The measured DC-link voltage is passed through a first-order low-pass filter and regulated by a PI controller to track its reference value. The grid-side converter is synchronized with the grid through a phase-locked loop and connected to the point of common coupling through an $RLC$ filter. Its $d$-axis current regulates the DC-link voltage and thus the active power exchange $P_\mathrm{UPS}$ with the grid, whereas the $q$-axis current regulates the reactive power $Q_\mathrm{UPS}$. Finally, the current-control dynamics are represented by first-order low-pass filters.

\section{Analysis of Data Center Load Impact}

\subsection{Power System Modeling} \label{sec:modeling}
We represent a power system with $n$ first-order differential equations and $m$ algebraic equations:
\begin{subequations}
\label{eq:system}
    \begin{align}
        \dot{\mathbf{x}} & = \mathbf{f}(\mathbf{x}, \mathbf{y}, \mathbf{u}), \label{eq:system_x} \\
        \mathbf{0}       & = \mathbf{g}(\mathbf{x}, \mathbf{y}, \mathbf{u}) ,\label{eq:system_0} \\
        \mathbf{z}       & = \mathbf{h}(\mathbf{x}, \mathbf{y}, \mathbf{u}) ,\label{eq:system_z}
    \end{align}
\end{subequations}
where $\mathbf{x} \in \mathbb{R}^n$ is the vector of state variables, $\mathbf{y} \in \mathbb{R}^m$ is the vector of algebraic variables, $\mathbf{u} \in \mathbb{R}^r$ is the vector of inputs, and $\mathbf{z} \in \mathbb{R}^o$ is the vector of outputs.

After the linearization of the nonlinear functions $\mathbf{f}$, $\mathbf{g}$, and $\mathbf{h}$ around the operating point $(\mathbf{x}^*,\mathbf{y}^*,\mathbf{u}^*)$, we obtain:
\begin{subequations}
    \begin{align}
        \Delta \dot{\mathbf{x}} & =
        \mathbf{A}
        \Delta \mathbf{x}  +
        \mathbf{B}
        \Delta \mathbf{u},  \label{eq:mat_ab} \\
        \Delta \mathbf{z}       & =
        \mathbf{C}
        \Delta \mathbf{x}  +
        \mathbf{D}
        \Delta \mathbf{u}, \label{eq:mat_cd}
    \end{align}
\end{subequations}
where $\mathbf{A} \in \mathbb{R}^{n \times n}$ is the state matrix, $\mathbf{B} \in \mathbb{R}^{n \times r}$ is the input matrix, $\mathbf{C} \in \mathbb{R}^{o \times n}$ is the output matrix, and $\mathbf{D} \in \mathbb{R}^{o \times r}$ is the feedforward matrix.
 
Assuming the IT load input is filtered and thus enters the system through a differential state, $\mathbf{D}=0$ and the resulting Transfer Function (TF) of the Single-Input Single-Output (SISO) system is:
\begin{equation}
\label{eq:transferfunction}
    G_S(s) = \frac{\Delta z(s)}{\Delta u(s)} =  
    \sum_{i=1}^n \frac{\mathbf{c} \boldsymbol{v}_i \boldsymbol{w}_i \mathbf{b}}{s-\lambda_i} 
    = \sum_{i=1}^n \frac{R_i}{s-\lambda_i} , 
\end{equation}
where $\mathbf{b}$ and $\mathbf{c}$ are the input and output vectors, respectively, $\boldsymbol{v}_i$ and $\boldsymbol{w}_i$ are the right and left eigenvectors of $\mathbf{A}$ associated with eigenvalue $\lambda_i = \sigma_i + j \omega_i$, while $R_i$ is the residue of the TF at $\lambda_i$. Hence:
\begin{itemize}
    \item The location of the cyclic load, represented in $\mathbf{b}$, plays a major role in determining its impact on the system.
    \item System components are affected differently by the cyclic load, as their location, represented by $\mathbf{c}$, determines how strongly the excited mode appears in their output.
\end{itemize}

Based on the principle of superposition, the steady-state response of a SISO system to the square-wave cyclic load in \eqref{eq:fourier_series} with 50\% duty cycle can be expressed as the sum of the responses to its individual harmonic components:
\begin{equation}
    \Delta z_{ss}(t) = \sum_{h=1}^{\infty}
    \Re \left\{\!
    G_S(jh\omega_f)
    \! A_h\, e^{j(h\omega_f t + \theta_h)} \right\} ,
    \label{eq:delta_z_ss_harmonics}
\end{equation}
where $A_h$ and $\theta_h$ are given by \eqref{eq:harmonic_amplitude} and \eqref{eq:harmonic_phase}, respectively. 
The amplitude contribution of the $h$-th harmonic to the output is:
\begin{equation}
    z^{(h)}_{ss} = \left|G_S(jh\omega_f)\right| A_h. 
    \label{eq:z_ss_perh_amp_exact}
\end{equation}

According to \eqref{eq:transferfunction} and \eqref{eq:z_ss_perh_amp_exact}, resonance occurs when $h\omega_f$ is close to $\omega_i = \omega_{n_i} \sqrt{1-\zeta_i^2} $, where $\omega_{n_i}$ is the undamped natural frequency, and $\zeta_i$ is the damping ratio of $\lambda_i$~\cite{kundur94}. As reported in~\cite{Zhou22}, the nonlinearities of the power system lead to small frequency deviations from the linear resonant frequency. These deviations are more noticeable at higher $P_f$ values and low-damping conditions. Additionally, \eqref{eq:delta_z_ss_harmonics} suggests that not only the fundamental component of the cyclic load can excite modes, but also the higher-order harmonics. However, as shown in \eqref{eq:harmonic_amplitude}, the higher-order harmonics have a lower impact compared to the fundamental since their amplitudes decrease with increasing harmonic order. Consequently, the resulting output oscillation also depends on the system gain at each harmonic frequency. Depending on the DTC training workload type, the fundamental, third, and fifth harmonic components may excite inter-area and local modes, while the others could excite sub-synchronous modes for large enough load fluctuations. The latter are beyond the scope of this study.

In the case of two identical and synchronized cyclic loads, such as two DTCs, the system response can be expressed as the sum of their individual SISO responses:
\begin{equation}
    \Delta z_{ss,2}(t) = 
    \sum_{h=1}^{\infty}
    \Re \left\{
    \left( G_{S}^1 + G_{S}^2 \right) 
    A_h\, e^{j(h\omega_f t + \theta_h)} 
    \right\} ,
    \label{eq:delta_z_ss_2}
\end{equation}
where $G_{S}^1$ and $G_{S}^2$ denote the individual TFs in \eqref{eq:transferfunction} associated with each DTC, while their explicit dependence on $jh\omega_f$ is omitted for brevity. The amplitude of the $h$-th harmonic is:
\begin{equation}
    z_{ss,2}^{(h)} =
    A_h \sqrt{
    \left| G_{S}^1\right|^2 +
    \left| G_{S}^2\right|^2 +
    2 \left| G_{S}^1\right| \left| G_{S}^2\right| 
    \cos\left( \Delta \phi_h \right)
    },
    \label{eq:delta_z_ss_2_mag}
\end{equation}
where $ \Delta \phi_h = \angle G_{S}^{1} - \angle G_{S}^{2} $ is the phase difference between the corresponding TFs at the frequency $h\omega_f$.

Eq.~\eqref{eq:delta_z_ss_2_mag} shows how the individual $h$-th harmonic responses produced by each DTC combine when both DTCs are present. For $\Delta\phi_h \in [-90^\circ,90^\circ]$, the cosine term in \eqref{eq:delta_z_ss_2_mag} is non-negative, and the responses reinforce each other, resulting in a larger harmonic amplitude than that with a single DTC described by \eqref{eq:z_ss_perh_amp_exact}. For $\Delta\phi_h \in (90^\circ,270^\circ)$, the responses partially cancel, while the total harmonic amplitude is smaller than \eqref{eq:z_ss_perh_amp_exact}~only if:
\begin{equation}
    \cos(\Delta\phi_h)
    <
    -\frac{\left|G_S^2\right|}
    {2\left|G_S^1\right|}.
\label{eq:two_dc_reduction_condition}
\end{equation}
For unsynchronized DTCs, a relative time offset $\Delta t$ adds $-h\omega_f\Delta t$ to $\Delta\phi_h$, while contributions from DTCs at different $\omega_f$ must be treated separately, regardless of their $P_f$.

\subsection{Stability Analysis of Periodic Solutions} \label{sec:Floquet}
The sustained response of the nonlinear Differential-Algebraic Equation (DAE) system \eqref{eq:system_x}--\eqref{eq:system_0} to the cyclic load is a periodic solution with the forcing period $T_{f}$, known as a forced limit cycle. This periodic solution (orbit) is characterized by:
\begin{subequations}
    \label{eq:periodic_solution}
    \begin{align}
        \mathbf{x}(t+T_f) &= \mathbf{x}(t), \label{eq:periodic_solution_x}\\
        \mathbf{y}(t+T_f) &= \mathbf{y}(t). \label{eq:periodic_solution_y}
    \end{align}
\end{subequations}
Its stability is assessed with Floquet theory, the periodic-orbit counterpart of eigenvalue analysis at equilibrium. 

Small perturbations around the periodic solution evolve according to the variational equation~\cite{Seydel10}:
\begin{equation}
    \dot{\boldsymbol{\Phi}}(t) = \mathbf{A}(t)\, \boldsymbol{\Phi}(t), \qquad \boldsymbol{\Phi}(0) = \mathbf{I},
    \label{eq:variational}
\end{equation}
where $\boldsymbol{\Phi}(t)$ is the state-transition matrix and $\mathbf{A}(t)$ is the $T_f$-periodic state matrix of \eqref{eq:mat_ab} evaluated along the orbit rather than at the equilibrium, with the algebraic variables eliminated through \eqref{eq:system_0}. Hard limits in the controls, such as the field-voltage limiter of excitation systems, make the model piecewise smooth. Whenever a limit engages along the orbit, $\boldsymbol{\Phi}$ is corrected with the corresponding saltation matrix~\cite{Hiskens00}.
    
The monodromy matrix $\mathbf{M}=\boldsymbol{\Phi}(T_f)$ maps an infinitesimal perturbation over one full period, and its eigenvalues $\mu_i$ are the Floquet multipliers of the orbit. For a constant state matrix $\mathbf{A}$, the multipliers reduce to $\mu_i = e^{\lambda_i T_f}$. Therefore, the stability conditions are:
\begin{equation}
    \Re \{ \lambda_i\} < 0 \Rightarrow \left|{e^{\lambda_iT_f}} \right| < 1, \quad i=1,\ldots,n.
\end{equation}

For a time-varying $\mathbf{A}(t)$, however, the eigenvalues of $\mathbf{A}(t)$ at frozen instants carry no stability information, and the multipliers must be computed from $\mathbf{M}$~\cite{Seydel10}. The orbit is stable if all multipliers lie inside the unit circle~\cite{Garcia22}, which thus plays the role of the left half-plane in the eigenvalue analysis of equilibrium points.

As $P_f$ increases, the manner in which a Floquet multiplier exits the unit circle classifies the instability~\cite{Kuznetsov04}. A real multiplier crossing $+1$ indicates a Saddle-Node (fold) Bifurcation (SNB) of the orbit: the stable limit cycle collides with an unstable one and both disappear. Beyond the bifurcation point, the trajectories may evolve toward another attractor, larger-amplitude oscillations, or collapse, depending on the global dynamics. Conversely, an exit from the unit circle at $-1$ or via a complex conjugate pair indicates a period-doubling (flip) or a Neimark-Sacker (torus) bifurcation, respectively. 

The multipliers capture instabilities of dynamic origin, but the differential-algebraic model \eqref{eq:system_x}--\eqref{eq:system_0} admits another instability mechanism. According to the implicit function theorem, the algebraic variables $\mathbf{y}$ can be locally expressed as functions of $\mathbf{x}$ only while the algebraic Jacobian $\mathbf{g}_{\mathbf{y}} = \partial \mathbf{g} / \partial \mathbf{y}$ is nonsingular~\cite{TVC98}. Let $\mathbf{f} \triangleq \mathbf{f}(\mathbf{x},\mathbf{y})$ and $\mathbf{g} \triangleq \mathbf{g}(\mathbf{x},\mathbf{y})$, with the input
$\mathbf{u}$ held at its nominal value. Among the equilibrium points of \eqref{eq:system}, the set
\begin{equation}
    \mathrm{SIB} = \left\{ (\mathbf{x},\mathbf{y}) \middle | \mathbf{f}=\mathbf{0}, \mathbf{g}=\mathbf{0}, \det \mathbf{g}_{\mathbf{y}} = 0 \right\} 
\end{equation}
corresponds to a Singularity-Induced Bifurcation (SIB), where an eigenvalue $\lambda_i$ diverges to infinity and stability is lost instantaneously~\cite{Venkatasubramanian95}. 
    
Along a forced limit cycle, the Impasse-Surface (IS) of the system \eqref{eq:system} consists of the set
\begin{equation}
    \mathrm{IS} = \left\{ (\mathbf{x},\mathbf{y})  \middle |  \mathbf{g}=\mathbf{0}, \det \mathbf{g}_{\mathbf{y}} = 0 \right\}. 
\end{equation}
This is a hypersurface where $\mathbf{g}_{\mathbf{y}}$ becomes singular~\cite{Song23}. When the system trajectories reach $\mathrm{IS}$, the periodic solution ceases to exist because the algebraic equations no longer admit a solution~\cite{Praprost96}. Since the Floquet multipliers may not yet have crossed the unit circle, this mechanism requires a separate detection criterion. To this end, we track $\min_{t \in [0,T_f)}\!\sigma_{\min}(\mathbf{g}_{\mathbf{y}}(t))$, the smallest singular value of $\mathbf{g}_{\mathbf{y}}$ minimized over one period, which approaches zero at the IS. 
    
Summarizing, a SNB of the periodic solution defines the maximum forcing amplitude that the system can sustain, analogous to loadability limits at SNBs of equilibrium points~\cite{TVC98}, beyond which instability occurs; encountering an IS of the DAE system produces an abrupt voltage collapse through a distinct mechanism~\cite{Song23}. Section~\ref{section:results_1} illustrates both mechanisms.

\section{Results}

The following small-signal analysis and dynamic simulations were performed using the RMS approximation based on a Python implementation and STEPSS~\cite{STEPPS14}. The first test system is used to assess the impact of a DTC on a small power grid under different levels of system strength and to illustrate the two instability mechanisms introduced in Section~\ref{sec:Floquet}. The simulations on the second system demonstrate the frequency-shift phenomenon in nonlinear systems, the effect of different square-wave duty cycles, and the impact of the third and fifth harmonics. The last test system is used to identify the DTC locations with the highest grid impact, as introduced in Section~\ref{sec:modeling}, and to analyze the interaction of DTCs running synchronized and unsynchronized jobs at different grid locations and forcing frequencies.

\subsection{Four-Bus Test System}
\label{section:results_1}

\begin{figure}[b!]
    \centering
    \vspace{-0.4cm}
    \includegraphics[width=0.95\columnwidth]{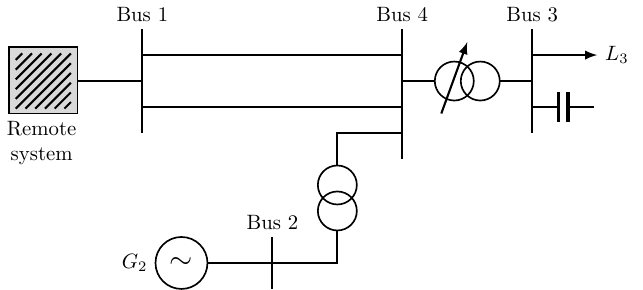}
    \vspace{-0.2cm}
    \caption{One-line diagram of the four-bus test system~\cite{TVC98}.}
    \label{fig:4bustestsystem}
\end{figure}

Fig.~\ref{fig:4bustestsystem} presents the one-line diagram of the considered 50-Hz, four-bus test system. The remote system is represented as a Thévenin equivalent, whose Short Circuit Capacity (SCC) is varied from 1000~MVA to 10000~MVA. Generator $G_2$ is a synchronous machine represented by a third-order model. The Automatic Voltage Regulator (AVR) and excitation system are represented by a first-order model, and the mechanical torque applied by the turbine is assumed constant. The system parameters, including lines, transformers, and shunt compensation, are taken from~\cite{TVC98} on a 100~MVA base.

The aggregate load $L_3$ at Bus~3 has a total active-power demand of 1500~MW and comprises two components: a voltage-dependent load consuming 1350~MW and 750~MVar, and a constant-power DTC load consuming 150~MW. The active-power component of the voltage-dependent load is modeled as a constant-current load, whereas its reactive-power component is modeled as a constant-impedance load. The DTC operates in Eco mode, with switch $S_1$ closed, as depicted in Fig.~\ref{fig:ups}. For this test system, the cooling demand associated with the induction machine is neglected; thus, the IT-load demand, and hence the total DTC demand, follows a square-wave profile with a 50\% duty cycle around the 150~MW operating point, similar to that shown in Fig.~\ref{fig:data_center_demand}(c).  

\begin{figure}[t!]
    \centering
    \includegraphics[width=0.975\columnwidth]{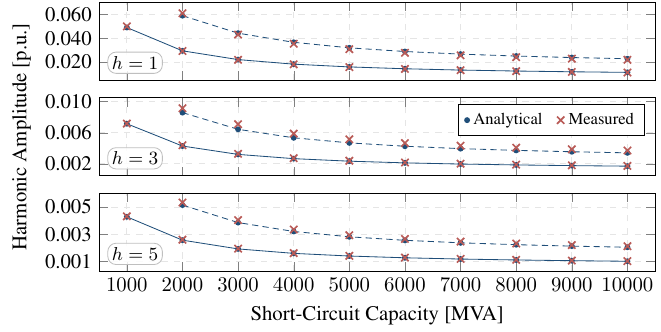}
    \vspace{-0.4cm}
    \caption{Analytically computed harmonic voltage amplitudes at Bus~2 using \eqref{eq:z_ss_perh_amp_exact} versus the values extracted from the time-domain simulation by FFT, for the first three odd harmonics ($h=1,3,5$, top to bottom) against different SCCs. Solid lines: $P_f=0.4$~p.u.; dashed lines: $P_f=0.8$~p.u. (the 1000~MVA case has no point as it lacks a bounded steady-state orbit).}
    \vspace{-0.4cm}
    \label{fig:4bus_harmonics}
\end{figure}

To assess the system response to sustained forced oscillations, the power fluctuation $P_f$ of the DTC load is varied at the frequency of the critical local mode corresponding to each SCC value. Then, the analytical harmonic amplitudes of the voltage magnitude at Bus~2, calculated using \eqref{eq:z_ss_perh_amp_exact}, are compared with those extracted from the time-domain simulation waveforms using a Fast Fourier Transform (FFT)~\cite{Oppenheim1997}. Fig.~\ref{fig:4bus_harmonics} presents this comparison for the voltage magnitude at Bus~2 at $P_f=0.4$ and $0.8$~p.u. with solid and dashed lines, respectively, considering the first three odd harmonics, since the even harmonics vanish for a 50\% duty cycle. At $P_f=0.8$~p.u., the 1000~MVA case is omitted, as the system no longer exhibits a bounded steady-state response at this disturbance level, as shown in the following simulations.

Across the entire SCC range, the analytically-derived system responses closely follow the FFT values. Moreover, the amplitudes of all three harmonics decrease progressively as SCC increases, reflecting the reduced voltage sensitivity of the stiffer systems, while the fundamental remains the dominant component throughout. For the 2000~MVA SCC case at $P_f=0.8$~p.u., the harmonic amplitudes are underestimated by approximately $3.3\%$. In this case, the field limiter becomes active during the forced response, introducing a saturation that is not represented in the linearized model. At $P_f=0.4$~p.u., the response remains closer to the small-signal regime, reducing the errors to approximately 2\% for all SCCs. 

\begin{figure}[t!]
    \centering
    \includegraphics[width=0.95\columnwidth]{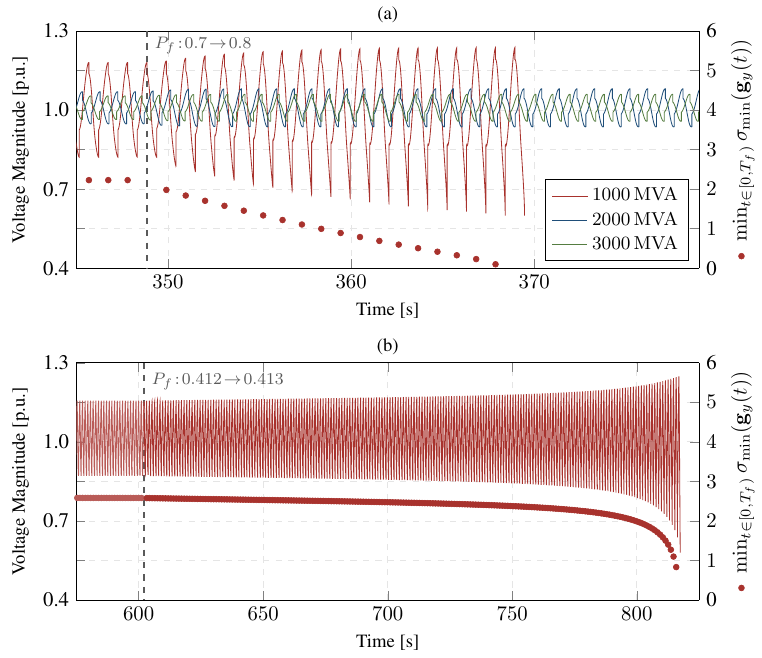}
    \vspace{-0.4cm}
    \caption{Voltage magnitude at Bus~2 (left axis) due to square-waveform load fluctuation at Bus~3 with increasing $P_f$, together with $\min\,\sigma_{\min}(\mathbf{g}_{\mathbf{y}})$ for the 1000~MVA SCC case (right axis), where $\omega_f$ equals the imaginary part of the critical local mode: (a) base system parameters for different SCCs; (b) reduced damping (10\% of initial) for 1000~MVA SCC.}
    \vspace{-0.4cm}
    \label{fig:4bus_voltage}
\end{figure}

Next, the analysis is restricted to the three lowest SCC values, namely, 3000, 2000, and 1000~MVA. These weak-grid cases are deliberately selected due to their greater sensitivity to the DTC forced oscillations. The forced oscillation is initiated at $t=1$~s, and $P_f$ is increased in 0.1~p.u. steps at approximately 50~s intervals until instability occurs. The voltage-magnitude responses at Bus~2 during the final increase in $P_f$ are shown in Fig.~\ref{fig:4bus_voltage}(a).

For the two higher-SCC cases, namely 2000 and 3000~MVA, the voltage settles into bounded periodic oscillations. In contrast, the weakest system becomes unstable during this final disturbance step. Specifically, increasing $P_f$ from 0.7 to 0.8~p.u. leads to growing voltage oscillations and, ultimately, instability, even though the corresponding local mode is small-signal stable with a damping ratio of 24.54\%. This result illustrates how sustained forced oscillations from DTCs can destabilize a nonlinear system that is small-signal stable.

\begin{figure}[b!]
    \centering
    \vspace{-0.4cm}
    \includegraphics[width=0.95\columnwidth]{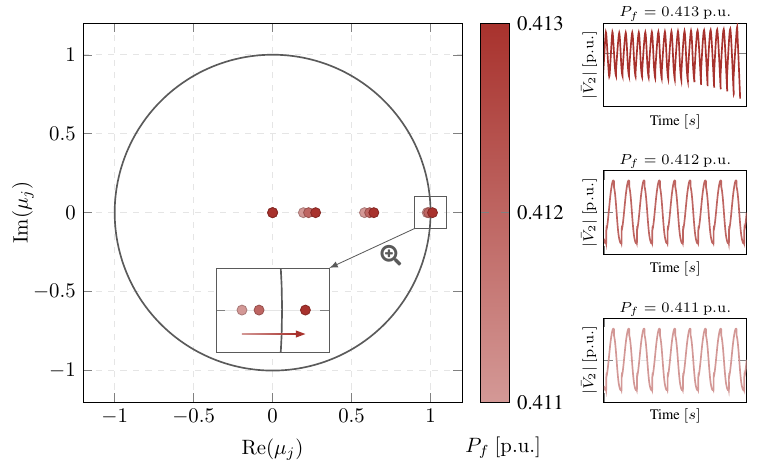}
    \vspace{-0.4cm}
    \caption{Floquet multipliers $\mu_j$ of the 1000~MVA SCC orbit under reduced damping for one representative window per $P_f$ level, with the corresponding Bus~2 voltage magnitudes shown in the insets.}
    \label{fig:4bus_floquet}
\end{figure}

As the disturbance magnitude is increased in steps from $P_f=0.1$~p.u., the 1000~MVA SCC case settles onto a stable periodic orbit at each level, and the dominant Floquet multiplier grows monotonically while remaining well inside the unit circle: its magnitude stays near 0.20 up to $P_f=0.3$~p.u., then rises to 0.809 at 0.7~p.u. In parallel, the smallest singular value of $\mathbf{g}_y$ over one period, $\min_{t\in[0,T_f)}\sigma_{\min}\!\left(\mathbf{g}_y(t)\right)$, decreases steadily over the same $P_f$ increments, from 3.39 at 0.1~p.u. to 2.23 at 0.7~p.u., indicating that the algebraic Jacobian $\mathbf{g}_y$ progressively approaches singularity even though each orbit remains stable. After the final increase to $P_f=0.8$~p.u., however, the trajectory becomes unstable before converging to a new periodic steady state. Strictly speaking, the Floquet multipliers are therefore not defined, since no periodic orbit is reached. Nevertheless, transient estimates remain informative; evaluated over successive forcing-period windows, the dominant multiplier stays well inside the unit circle roughly at 0.74--0.81 magnitude with no tendency to approach unity. In contrast, $\min_{t\in[0,T_f)}\sigma_{\min}\!\left(\mathbf{g}_y(t)\right)$, shown on the right-hand axis of Fig.~\ref{fig:4bus_voltage}(a), decreases monotonically from approximately 1.98 at the onset of this interval to about 0.10 roughly 1.6~s before the simulation terminates. This behavior indicates IS instability introduced in Section~\ref{sec:Floquet}, where the periodic orbit ceases to exist as $\mathbf{g}_y$ becomes singular along the trajectory, rather than a Floquet multiplier crossing the unit~circle.

In order to isolate a genuine SNB of the periodic orbit, as opposed to the IS mechanism observed above, the simulation of the 1000~MVA SCC case is repeated with the damping coefficient of $G_2$ reduced to 10\% of its original value. Accordingly, $P_f$ is swept in fine increments near 0.41~p.u., namely, 0.411, 0.412, and 0.413~p.u., with the forcing frequency fixed at the critical eigenvalue frequency. Fig.~\ref{fig:4bus_voltage}(b) shows the corresponding Bus~2 voltage magnitude over the last two holds, together with $\min_{t\in[0,T_f)}\sigma_{\min}\!\left(\mathbf{g}_y(t)\right)$ on the right axis, while Fig.~\ref{fig:4bus_floquet} shows the associated Floquet multipliers.

As $P_f$ increases, the dominant multiplier, located on the real axis, moves progressively to the right, from 0.979 to 0.988 and finally to 1.012, exiting the unit circle through the real axis between the last two levels, indicating a SNB. Unlike in the previous IS case, $\min_{t\in[0,T_f)}\sigma_{\min}\!\left(\mathbf{g}_y(t)\right)$ remains significantly larger than zero throughout the sweep, taking values of 2.66, 2.58, and 2.45 at the three representative windows, respectively. This confirms that the instability is of dynamic rather than algebraic origin. As discussed in Section~\ref{sec:Floquet}, the SNB can be associated with a growing oscillatory response, as seen in the time-domain voltage-magnitude response at Bus~2 in Fig.~\ref{fig:4bus_voltage}(b), which grows from cycle to cycle following the last $P_f$ step until the simulation terminates at $t\approx817$~s.

Interestingly, shifting the forcing frequency away from the critical frequency allows the system to sustain substantially larger forced-oscillation magnitudes, namely 0.746~p.u. and 1.363~p.u. for forcing frequencies 10\% below and above the critical frequency, respectively. Moreover, the mechanisms driving the instability differ between the two cases: in the former, the instability is caused by a SNB, whereas in the latter, it results from an IS. These findings highlight the sensitivity of the system to both the magnitude and frequency of the forced oscillation and further illustrate how the underlying instability mechanism strongly depends on system operating conditions. The corresponding results are omitted due to space limitations.

\subsection{Two-Area Test System}
\label{section:results_2}
\begin{figure}[b!]
    \centering
    \vspace{-0.4cm}
    \includegraphics[width=0.95\columnwidth]{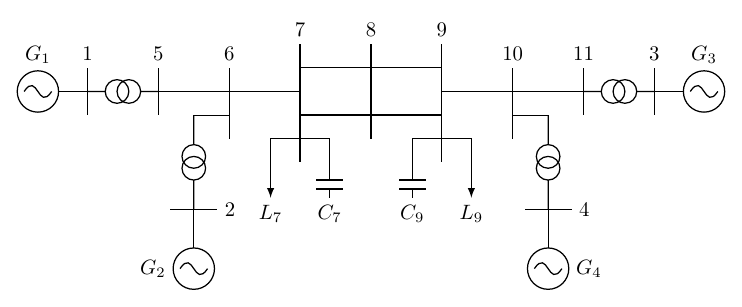}
    \vspace{-0.4cm}
    \caption{One-line diagram of the two-area test system~\cite{kundur94}.}
    \label{fig:kundur_system}
\end{figure}

The second test system consists of two similar areas connected by a weak interconnection at 230~kV, as shown in Fig.~\ref{fig:kundur_system}. This 60-Hz, 100~MVA base test system, prone to local and inter-area oscillations~\cite{kundur94}, consists of four synchronous generators $G_1$-$G_4$, each represented by a sixth-order model (round-rotor machines). The generators are equipped with thyristor-based excitation systems with high transient gain and a Power System Stabilizer (PSS) consisting of a washout filter and two cascaded lead/lag blocks. To analyze a scenario with relatively low system damping, the PSS gains in all generators are fixed to 10 p.u./p.u. to achieve $\zeta \leq 5\%$ for the most critical electromechanical mode. The loads $L_7$ and $L_9$ are modeled as constant current for active power and constant impedance for reactive power, with shunt capacitors $C_7$ and $C_9$. The system parameters and operating point are specified in~\cite{kundur94}.

\begin{figure}[b!]
    \centering
     \vspace{-0.4cm}
    \includegraphics[width=0.95\columnwidth]{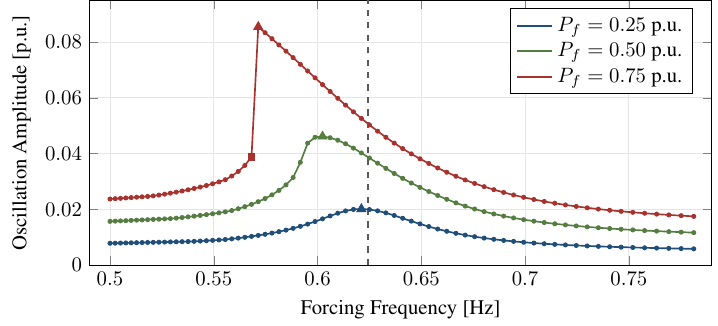}
    \vspace{-0.4cm}
    \caption{Oscillation amplitude in Bus~9 voltage magnitude against the forcing frequency of the DTC load for three $P_f$ values. The dashed line marks the critical eigenvalue frequency $f_c=0.6244$~Hz. Triangles mark the frequency leading to maximum oscillation amplitude, while the square marks the SNB of the forced periodic orbit at $P_f=0.75$~p.u.}
    \label{fig:kundur_frequencies}
\end{figure}

Since there are two candidate load buses (Buses 7 and 9), the DTC is placed at the bus where its system impact is the highest. Hence, the DTC is placed at Bus~9, where it represents 10\% of the total active power demand at that bus. With the DTC connected, the critical mode equals $-0.1634 \pm j3.9233$, with $\zeta =4.161\%$ and a frequency of $f_c=0.6244$~Hz.

To account for frequency shifts~\cite{Zhou22}, the system is simulated with different DTC forced oscillation magnitudes $P_f$ and frequencies $f_f$ around $f_c$. For the following analyses, the DTC operates in UPS mode, with switch $S_2$ closed as shown in Fig.~\ref{fig:ups}. The resulting forced oscillations in the voltage magnitude at Bus~9 are shown in Fig.~\ref{fig:kundur_frequencies}. Due to the frequency shift phenomenon, the amplitude peaks do not occur at $f_c$ but move progressively farther from it as the forcing magnitude grows: from $0.6211$~Hz at 0.25~p.u. to $0.6024$~Hz at 0.50~p.u. and $0.5714$~Hz at 0.75~p.u. Moreover, at $P_f=0.75$~p.u., the forced oscillation amplitude experiences an abrupt change, known as a \textit{jump phenomenon} of nonlinear systems, indicated by the red square and triangle points in Fig.~\ref{fig:kundur_frequencies}. This indicates another SNB caused by small increments in $f_f$ for a fixed $P_f$. In contrast to the SNB observed in Section~\ref{section:results_1}, a new stable orbit is reached: the small-amplitude limit cycle shifts to the large-amplitude solution as the only attractor.

\begin{figure}[t!]
    \centering
    \includegraphics[width=0.95\columnwidth]{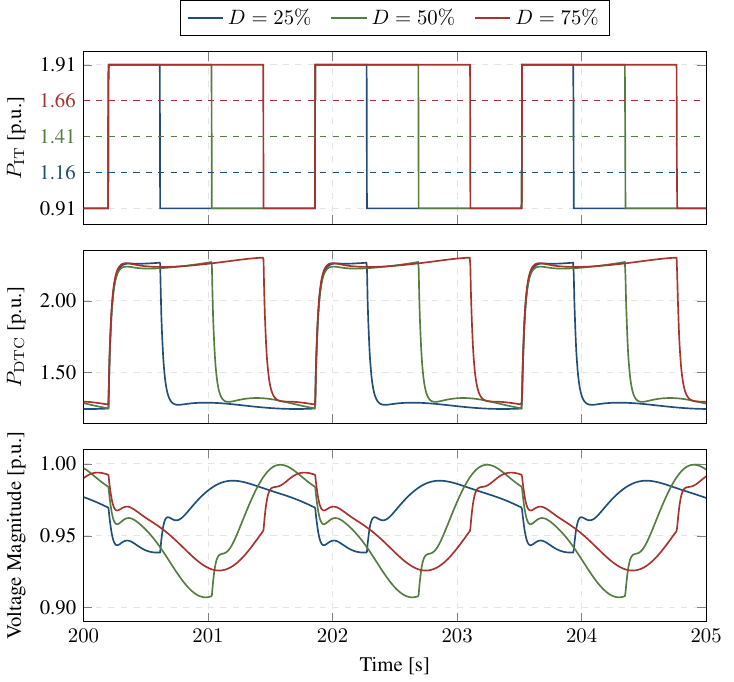}
    \vspace{-0.4cm}
    \caption{Impact of different DTC duty cycles $D$ on the voltage magnitude at Bus~9 for a DTC connected at Bus~9 with $P_f=0.5$~p.u.: IT-load demand $P_{\mathrm{IT}}$ with dashed lines indicating the corresponding mean values (upper panel), total DTC demand $P_{\mathrm{DTC}}$ in UPS mode, including the cooling load shown in Fig.~\ref{fig:ups} (middle panel), and the voltage magnitude at Bus~9 (lower panel).}
    \label{fig:kundur_duty_cycles}
    \vspace{-0.4cm}
\end{figure}

Because DTC load fluctuations can vary considerably with AI workload, the impact of different duty cycles $D$ is also investigated. In the following analysis, the DTC is simulated with $P_f=0.5$~p.u. at the resonant forcing frequency of 0.6024~Hz, which yields the peak voltage deviations for the considered $P_f$, as discussed above. As the DTC operates in UPS mode, the total DTC demand $P_\mathrm{DTC}$ does not retain the ideal square-wave profile of the IT load $P_\mathrm{IT}$, as shown in the middle and upper panels of Fig.~\ref{fig:kundur_duty_cycles}, respectively. Moreover, for duty cycles different from 50\%, the mean IT-load demand is shifted according to \eqref{eq:fourier_series}, as indicated by the dashed lines in the upper panel. The corresponding voltage-magnitude responses at Bus~9 are shown in the lower panel. For $D=50\%$, the equal forcing intervals allow the voltage magnitude to complete a swing during each half-cycle. In contrast, the shorter forcing intervals for $D=75\%$ and 25\% are insufficient for a complete voltage swing, resulting in narrow spikes with signs opposite to the corresponding power deviations: positive voltage spikes during the low-power intervals for $D=75\%$ and negative spikes during the high-power intervals for $D=25\%$. These spikes may pose risks to voltage-sensitive devices. 

\begin{figure}[b!]
    \centering
    \vspace{-0.4cm}
    \includegraphics[width=0.95\columnwidth]{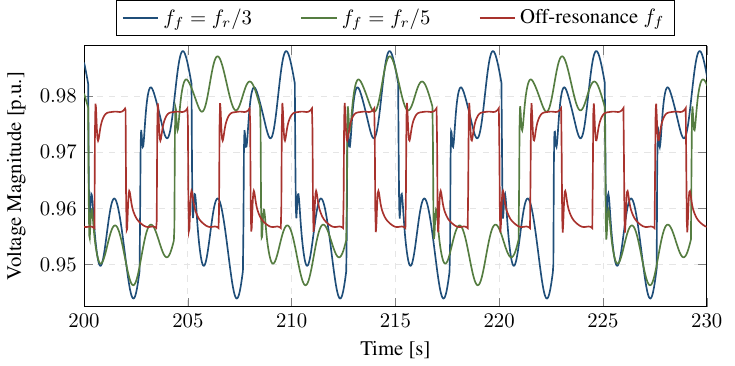}
    \vspace{-0.4cm}
    \caption{Impact of different forced oscillation frequencies on voltage magnitude at Bus~9 for $P_f=0.5$~p.u.}
    \label{fig:kundur_harmonics}
\end{figure}

The second simulation for this test system evaluates the impact of the third and fifth harmonic components of the square-wave load represented by \eqref{eq:fourier_series} with $D=50\%$. To achieve this, we assumed DTC load fluctuations with frequencies of 0.2008~Hz and 0.1205~Hz, such that the respective third and fifth harmonics are approximately equal to the resonant frequency of $f_r=0.6024$~Hz. For comparison purposes, we also evaluated an additional off-resonance frequency ($0.33$~Hz), which does not match any system mode frequency. 

Fig.~\ref{fig:kundur_harmonics} presents the simulation results, showing that DTC load fluctuations can induce large system oscillations when one of their harmonic frequencies coincides with $f_r$. As expected from \eqref{eq:z_ss_perh_amp_exact}, the amplitudes are smaller than for $f_f=f_r$.

\subsection{Longitudinal Test System}

\begin{figure}[b!]
    \centering
    \vspace{-0.4cm}
    \includegraphics[width=0.95\columnwidth]{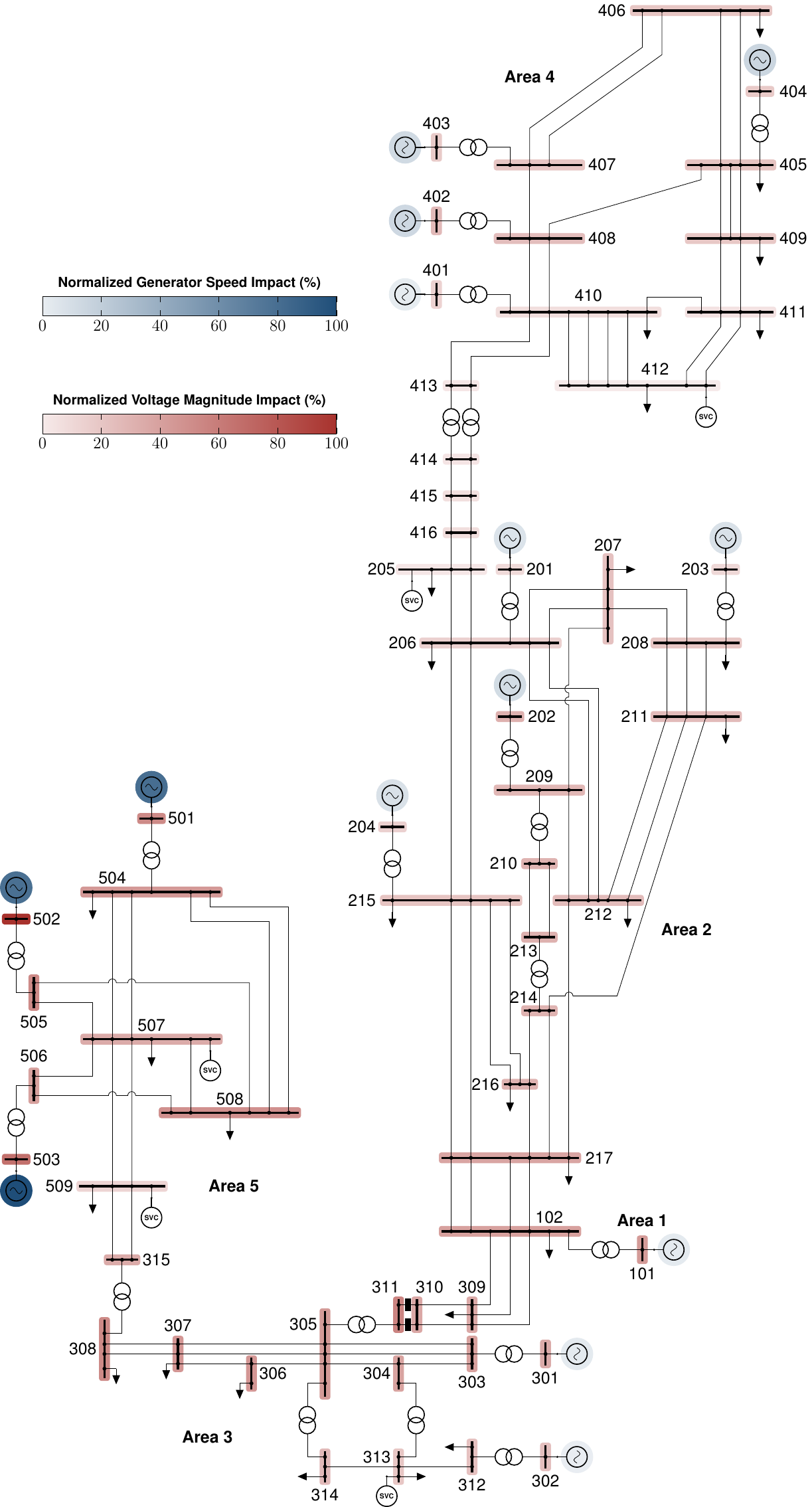}
    \vspace{-0.3cm}
    \caption{Longitudinal system diagram with normalized TF \eqref{eq:transferfunction} magnitudes for generator speeds (blue) and voltage magnitudes (red) for a DTC at Bus~508.}
    \label{fig:australian_system_heatmap}
\end{figure}

The third test system corresponds to the longitudinal power system illustrated in Fig.~\ref{fig:australian_system_heatmap}. This 50-Hz system is divided into five areas. There are fourteen synchronous generators represented by sixth-order (round-rotor machines) and fifth-order (salient-pole machines) models, including AC1A and AC4A excitation system models~\cite{IEEE421516}. The AVRs of these excitation systems are equipped with a PSS consisting of a washout filter and two cascaded lead/lag blocks. All loads are modeled as constant current for active power and constant impedance for reactive power. Additionally, there are five Static Var Compensators (SVCs) for local voltage control. The system parameters on a 100-MVA base are provided in~\cite{Gibbard15}. 

\begin{figure}[t!]
    \centering
    \includegraphics[width=0.995\columnwidth]{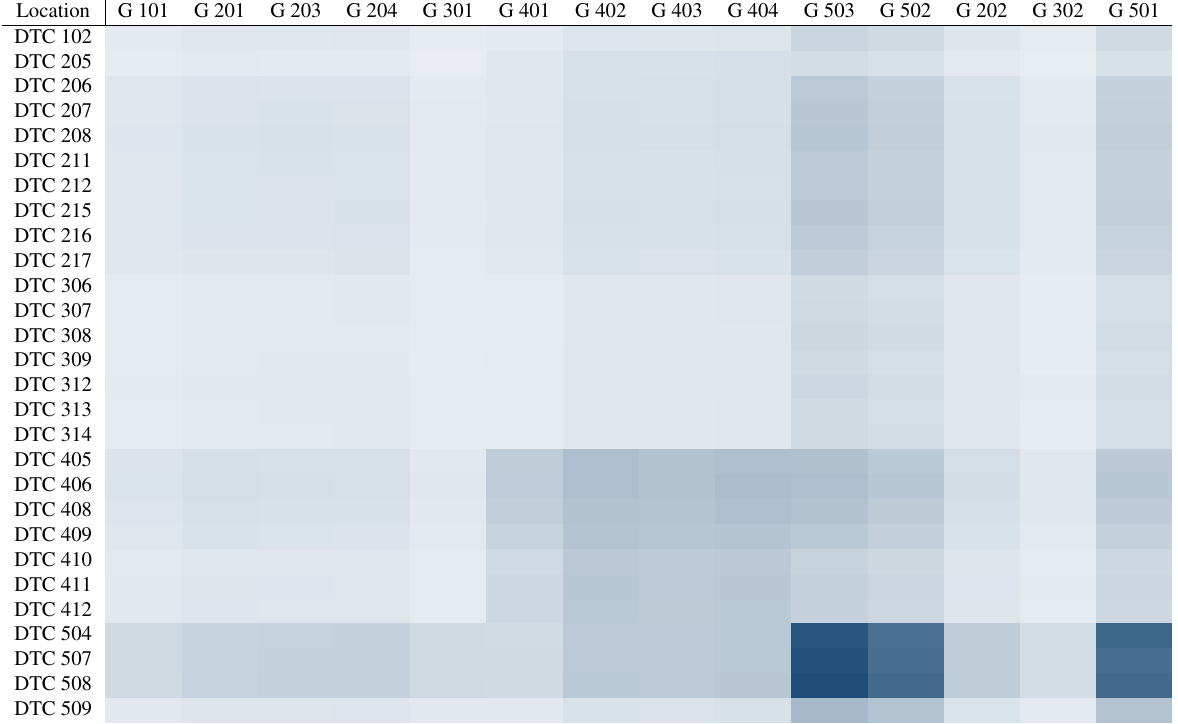}
    \vspace{-0.7cm}
    \caption{Generator speed heatmap for the longitudinal system based on the magnitude response of the TFs \eqref{eq:transferfunction} on a 100-MVA base.}
    \label{fig:australian_system_heatmap_G}
    \vspace{-0.5cm}
\end{figure}

The small-signal analysis reveals that the two most critical electromechanical modes are $-0.382 \pm j3.592$ ($ \zeta =10.57\%$) and $-0.344 \pm j2.768 $ ($ \zeta =12.33\%$), respectively. For each potential DTC location, the state-space matrices of the linearized model are obtained, and the magnitude response of the TF at the forcing frequency $f_f$ is computed from \eqref{eq:transferfunction} for the selected input (DTC demand at the location of interest) and desired output. The TF output can be, for example, a voltage magnitude or a machine speed. Other physical quantities can be defined as system outputs as long as they can be represented with the output matrix $\mathbf{C}$. Intuitively, the DTC locations with the highest impact on the grid are the ones that yield the largest magnitude response of the TFs, evaluated at the DTC load fluctuation frequency that matches the critical mode frequency.

Fig.~\ref{fig:australian_system_heatmap_G} presents the heatmap of DTC locations and corresponding impact on the fourteen generator speeds based on the TF responses. The most critical DTC locations and the most impacted generators can be identified by the darkest rows and columns, respectively: DTCs at Buses 504, 507, and 508 create the greatest speed oscillations at generators 501, 502, and 503.

\begin{table}[b!]
    \vspace{-0.5cm}
    \caption{Critical DTC locations -- Impact on voltage magnitudes.}
    \vspace{-0.2cm}
    \centering
    \begin{tabular}{|c|c|c|c|c|}
    \hline
     Location & V501 & V502 & V503 &V311  \\ \hline
     Bus 504 & 0.027265	& 0.035472 & 0.024194  & 0.022109 \\ \hline
     Bus 507 & 0.021174 & 0.037915  & 0.026396 & 0.022680 \\ \hline
     Bus 508 & 0.022072	& 0.038873	& 0.027001 & 0.023123 \\ \hline 
    \end{tabular}
    \label{tab:australian_test_system_ssresponse_voltages}
\end{table}

The same procedure can be carried out for other selected outputs. Table~\ref{tab:australian_test_system_ssresponse_voltages} presents the TF \eqref{eq:transferfunction} magnitude responses for the most impactful DTC locations and most impacted buses in terms of voltage magnitudes, expressed in per unit on a 100-MVA base. Placing the DTC at Buses 504, 507, or 508 in Area 5 results in the most stressed grid conditions, with the most impacted buses in Area 5, specifically the voltage magnitude at Bus 502. Furthermore, Bus 311 in Area 3 is also affected by the DTC load fluctuation in Area 5. These results are in line with the theoretical expectations, as buses within Area 5 and Area 3 are electrically closer.

To further illustrate the most impacted buses and generators by DTCs, the magnitudes of the normalized TF response of generator speeds and voltage magnitudes are indicated using a heatmap in the one-line diagram of Fig.~\ref{fig:australian_system_heatmap}, for the DTC placed at the most critical location, i.e., at Bus 508. This location is identified based on both the largest TF magnitude at any single output and the largest TF magnitude averaged over all outputs. Moreover, additional simulations show that the maximum $P_f$ yielding bounded oscillations is 1.61, 2.17, and 0.62~p.u. on a 100~MVA base for DTCs at Buses 508, 507, and 504, i.e.\ 31.3\%, 32.3\%, and 32.8\% of the total active power at the point of connection, respectively. Among these locations, the ordering mirrors the TF-based ranking: the more critical the location, the smaller the relative load fluctuation it can sustain.

\begin{figure}[b!]
    \centering
    \vspace{-0.4cm}
    \includegraphics[width=0.95\columnwidth]{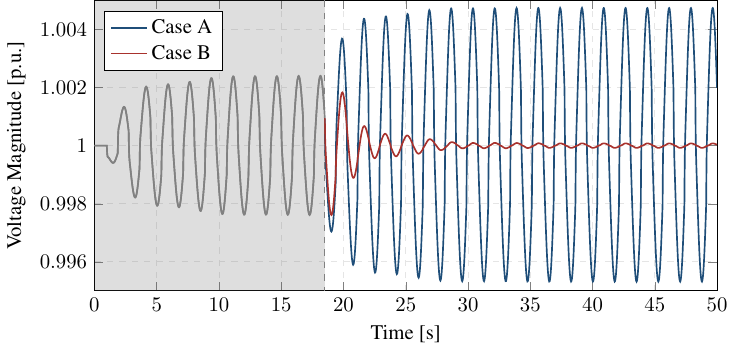}
    \vspace{-0.4cm}
    \caption{Impact of DTC synchronization on voltage magnitude at Bus~502, with the second DTC starting at 18.5~s (Case~A) and 19.375~s~(Case~B).}
    \label{fig:CaseA_CaseB_V502}
\end{figure}

\begin{figure}[t!]
    \centering
    \includegraphics[width=0.95\columnwidth]{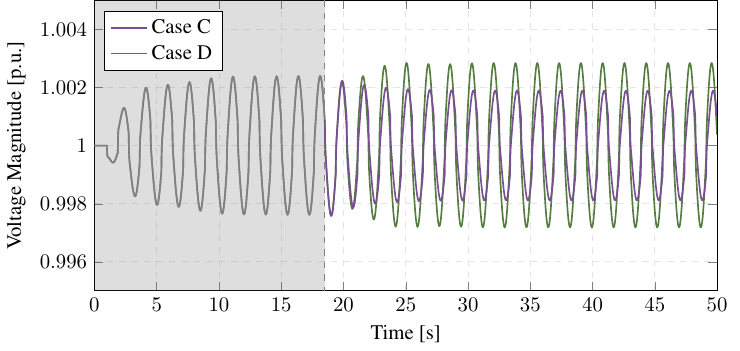}
    \vspace{-0.4cm}
    \caption{Impact of two DTCs oscillating at critical and less critical buses on voltage magnitude at Bus~502. The second DTC starts oscillating at $18.5$~s.}
    \label{fig:CaseC_CaseD_V502}
    \vspace{-0.4cm}
\end{figure}

\begin{figure}[b!]
    \centering
    \vspace{-0.40cm}
    \includegraphics[width=0.95\columnwidth]{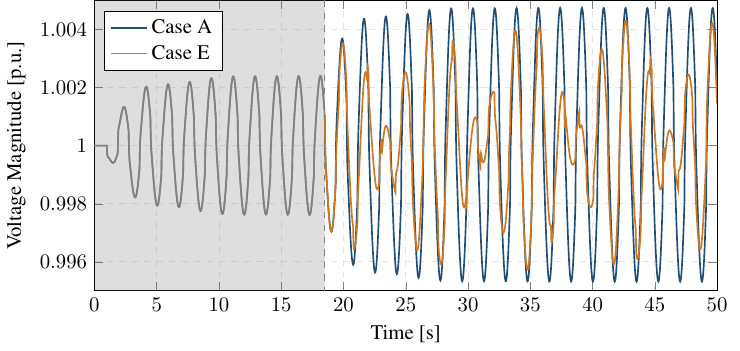}
    \vspace{-0.4cm}
    \caption{Impact of two DTCs oscillating at different frequencies on voltage magnitude at Bus 502. The second DTC starts oscillating at $18.5$~s.}
    \label{fig:CaseA_CaseE_V502}
\end{figure}

To assess the impact of multiple DTCs, simulations are conducted with two DTCs in the system. Five cases are considered, which differ in terms of DTC locations and oscillation frequencies, as summarized in Table~\ref{tab:Longitudinal_cases}. Each DTC adds a load equal to 5\% of the initial demand at its corresponding bus and operates in UPS mode with $P_f=0.05$~p.u. on the 100~MVA base. In all cases, the first DTC is located at Bus~508 and begins to fluctuate at 1~s, while the second DTC starts at 18.5~s (19.375~s in Case~B). The shaded region in Figs.~\ref{fig:CaseA_CaseB_V502}--\ref{fig:CaseA_CaseE_V502} thus shows the single-DTC response of the Bus~508 DTC, against which the two-DTC response is compared.

\begin{table}[t]
    \centering
    \caption{Considered cases for the oscillating DTCs.}
    \vspace{-0.2cm}
    \label{tab:Longitudinal_cases}
    \resizebox{0.95\columnwidth}{!}{\begin{tabular}{c c c c}
        \toprule
        \textbf{Case} & \textbf{DTC Locations} & \textbf{Synchronization} & \textbf{Frequency} \\
        \midrule
        A & Bus 508 \& 507 & Synchronized & Critical mode \\
        B & Bus 508 \& 507 & Unsynchronized & Critical mode \\
        C & Bus 508 \& 208 & Synchronized & Critical mode \\
        D & Bus 508 \& 405 & Synchronized & Critical mode \\
        E & Bus 508 \& 507 & -- & Two critical modes \\
        \bottomrule
    \end{tabular}}
    \vspace{-0.5cm}
\end{table}

Fig.~\ref{fig:CaseA_CaseB_V502} presents the voltage magnitude at the most affected Bus~502 for Cases~A and~B. The shaded area in the figure indicates that, during this time, only one DTC exhibits fluctuating power. Once the load fluctuation of the second DTC starts, the voltage variation is larger in Case~A (in phase) than in Case~B (out of phase). This behavior follows from the TFs \eqref{eq:transferfunction} defined from each DTC location to the voltage magnitude at Bus~502, evaluated at the fundamental (forcing) frequency which dominates the system response. Specifically, both DTCs are in the same area, and their TFs have nearly equal magnitudes and phases, so the phase difference $\Delta\phi_h$ between their contributions \eqref{eq:delta_z_ss_2_mag} is defined by the start offset $\Delta t$: it equals zero in Case~A, whereas the half-cycle offset of Case~B adds $180^\circ$ for every odd harmonic. Consequently, $\cos(\Delta\phi_h) \approx 1$ in Case~A, where the two contributions add and the oscillation approximately doubles, whereas $\cos(\Delta\phi_h) \approx -1$ in Case~B, where they oppose and, with \eqref{eq:two_dc_reduction_condition} satisfied, the amplitude falls to about 3\% of the single-DTC response. Hence, synchronized oscillatory behavior between two DTCs at critical locations in the same area negatively affects the voltage, whereas out-of-phase AI workloads can reduce the magnitude of the forced oscillations. These results suggest that scheduling the execution of AI workloads across DTCs could help reduce their impact on the power grid.

The effect of the second DTC location is shown in Fig.~\ref{fig:CaseC_CaseD_V502}, which compares Bus 502 voltage magnitudes for Cases C and D, with the first DTC at the most critical location and the second at a less critical location in Area 2 and Area 4, respectively. In particular, Case~C leads to smaller oscillations than Case~D. Applying the same TF-based analysis, both DTCs now start in phase, so that $\Delta\phi_h$ is determined solely by the TF phases. Specifically, the TFs \eqref{eq:transferfunction} of the DTCs at Buses~508 and 208 (Case C) differ in phase by approximately $152^\circ$; thus, their contributions partially cancel according to \eqref{eq:delta_z_ss_2_mag}. Moreover, as \eqref{eq:two_dc_reduction_condition} is also satisfied, the resulting amplitude decreases compared to the single-DTC response. Conversely, for DTCs at Buses~508 and 405 (Case D), the phase difference is approximately $57^\circ$, causing their responses to reinforce each other and resulting in larger voltage oscillations than the single-DTC response. This effect is output-dependent, with Cases~C and D exhibiting different spatial patterns of constructive and destructive interaction across the system. 

Finally, Fig.~\ref{fig:CaseA_CaseE_V502} compares the impact of the two DTC oscillation frequencies on the voltage magnitude at Bus~502 for Cases A and E. In particular, oscillations at the two most critical modal frequencies have a smaller impact than when both DTCs oscillate at the most critical frequency.

\section{Conclusion}

This paper investigated forced oscillations induced by AI workloads using analytical approaches and time-domain simulations. We found that large load fluctuations of DTCs could drive a small-signal stable system outside its region of attraction, leading to unbounded oscillations. The Floquet multipliers of the forced periodic response detect the SNB of the oscillation, whereas the IS, at which the multipliers give no warning, is detected by the minimum singular value of the algebraic Jacobian. For bounded oscillations, the derived equations enable harmonic analysis of the forced response using a linearized model, without requiring time-domain simulations.

Due to the nonlinear nature of the power system, the peak amplitude shifts progressively away from the modal frequency as fluctuations grow, so the worst-case forcing frequency is not the modal one. In the largest fluctuation magnitude considered, the \textit{jump phenomenon} is observed. Moreover, among the tested DTC load fluctuation duty cycles, we found that a 50\% duty cycle results in the largest amplitude of forced oscillations, compared to 25\% or 75\% duty cycles, while harmonic components beyond the fundamental frequency of the DTC can also result in large forced oscillations.

Finally, the magnitude response of the TFs can be used to identify critical DTC locations and regions of greatest impact, while their phase determines how multiple DTCs interact: synchronized DTCs at critical buses in the same area can nearly double the oscillation magnitude, a half-cycle offset can nearly cancel it, and for DTCs in different areas, the TF phase difference determines whether contributions reinforce or partially cancel. This suggests opportunities for AI workload scheduling as a practical mitigation measure.

\bibliographystyle{IEEEtran}
\typeout{}
\bibliography{ref.bib}

@book{Gibbard15,
  title     = {Small-Signal Stability, Control and Dynamic Performance of Power Systems},
  isbn      = {9781925261035},
  publisher = {The University Of Adelaide Press},
  author    = {M J Gibbard and P Pourbeik and Vowles, D J},
  year      = {2015}
}

@techreport{Ryan25,
  author      = {Quint, Ryan and Zhao, Jiecheng and Thomas, Kyle},
  title       = {An Assessment of Large Load Interconnection Risks in the Western Interconnection},
  institution = {WECC},
  address     = {Salt Lake City, UT, USA},
  month       = feb,
  year        = {2025},
  type        = {Technical Report},
}

@article{Pinneilo71,
  author={Pinneilo, J. A. and Van Ness, J. E.},
  journal = {IEEE Trans. Power App. Syst.},
  title   = {Dynamic Response of a Large Power System to a Cyclic Load Produced by a Nuclear Accelerator},
  year    = {1971},
  volume  = {PAS-90},
  number  = {4},
  pages   = {1856-1862},
  doi     = {10.1109/TPAS.1971.293180}
}

@article{Rao88,
  author  = {K. R. Rao and L. Jenkins},
  journal = {IEEE Trans. Power Syst.},
  title   = {Studies on power systems that are subjected to cyclic loads},
  year    = {1988},
  volume  = {3},
  number  = {1},
  pages   = {31-37},
  doi     = {10.1109/59.43178}
}

@techreport{Shehabi24,
  author      = {Shehabi, A. and Smith, S. J. and Hubbard, A. and Newkirk, A. and Lei, N. and Siddik, M. A. B. and Holecek, B. and Koomey, J. and Masanet, E. and Sartor, D.},
  title       = {2024 {United} {States} Data Center Energy Usage Report},
  institution = {Lawrence Berkeley National Laboratory},
  address     = {Berkeley, California},
  year        = {2024},
  number      = {LBNL-2001637},
  type        = {Technical Report}
}

@techreport{NERC25,
  author      = {{North American Electric Reliability Corporation (NERC)}},
  title       = {Characteristics and Risks of Emerging Large Loads},
  institution = {NERC},
  type        = {White Paper},
  month       = jul,
  year        = {2025},
  note        = {Large Loads Task Force}
}

@techreport{iea24,
  author      = {Çam, E. and Hungerford, Z. and Schoch, N. and Pinto, F. and Yáñez de León, C. D.},
  title       = {Electricity 2024: Analysis and Forecast to 2026},
  institution = {International Energy Agency (IEA)},
  year        = {2024}
}

@techreport{EPRI24,
  author = {{Electric Power Research Institute (EPRI)}},
  title       = {Powering Intelligence: Analyzing Artificial Intelligence and Data Center Energy Consumption},
  institution = {EPRI},
  type        = {White Paper},
  year        = {2024}
}

@article{Ye17,
  author  = {Ye, H. and Liu, Y. and Zhang, P. and Du, Z.},
  title   = {Analysis and Detection of Forced Oscillation in Power System},
  journal = {IEEE Trans. Power Syst.},
  volume  = {32},
  number  = {2},
  pages   = {1149--1160},
  month   = mar,
  year    = {2017},
  doi     = {10.1109/TPWRS.2016.2580710}
}

@article{Cai25,
  author  = {Cai, Y. and Pierrou, G. and Wang, X. and Joos, G.},
  title   = {A Data-Driven Forced Oscillation Locating Method for Power Systems with Inverter-Based Resources},
  journal = {IEEE Trans. Power Syst.},
  volume  = {41},
  number  = {2},
  pages   = {1192--1203},
  year    = {2026},
  doi     = {10.1109/TPWRS.2025.3600646}
}

@techreport{Chen23,
  author      = {Chen, L. and Trudnowski, D. and Dosiek, L. and Kamalasadan, S. and Xu, Y. and Wan, X.},
  title       = {Forced Oscillations in Power Systems ({TR} 110)},
  institution = {IEEE PES Task Force on Oscillation Source Location},
  month       = may,
  year        = {2023},
  doi         = {10.17023/k6ff-rr85}
}

@techreport{Chala25,
  author      = {B. Chalamala and others},
  title       = {Data Center Growth and Grid Readiness ({TR} 131)},
  institution = {IEEE PES Task Force on Data Center Growth and Grid Readiness},
  year        = {2025},
  month       = {May},
  doi         = {10.17023/w4wy-s557},
  type        = {Technical Report}
}

@article{Smolleck91,
  author  = {H. A. Smolleck and S. J. Ranade and N. R. Prasad and R. O. Velasco},
  journal = {IEEE Trans. Power Del.},
  title   = {Effects of pulsed-power loads upon an electric power grid},
  year    = {1991},
  volume  = {6},
  number  = {4},
  pages   = {1629-1640},
  doi     = {10.1109/61.97702}
}

@article{Kez20,
  author  = {D. A. Kez and A. M. Foley and S. M. Muyeen and D. J. Morrow},
  journal = {IEEE Access},
  title   = {Manipulation of Static and Dynamic Data Center Power Responses to Support Grid Operations},
  year    = {2020},
  volume  = {8},
  pages   = {182078-182091},
  doi     = {10.1109/ACCESS.2020.3028548}
}

@article{Sun22,
  author   = {Sun, Jian and Mihret, Melaku and Cespedes, Mauricio and Wong, David and Kauffman, Mike},
  journal  = {CSEE J. Power Energy Syst.},
  title    = {Data Center Power System Stability — {Part II}: System Modeling and Analysis},
  year     = {2022},
  volume   = {8},
  number   = {2},
  pages    = {420-438},
  doi      = {10.17775/CSEEJPES.2021.02020}
}

@book{kundur94,
  author    = {Prabha Kundur},
  title     = {Power System Stability and Control},
  publisher = {The EPRI Power System Engineering Series, McGraw-Hill},
  year      = {1994}
}

@book{TVC98,
  title     = {Voltage Stability of Electric Power Systems},
  publisher = {Springer},
  author    = {T {Van Cutsem} and C Vournas},
  year      = {1998}
}

@ARTICLE{STEPPS14,
  author={Aristidou, Petros and Fabozzi, Davide and Van Cutsem, Thierry},
  journal={IEEE Trans. Parallel Distrib. Syst.}, 
  title={Dynamic Simulation of Large-Scale Power Systems Using a Parallel {S}chur-Complement-Based Decomposition Method}, 
  year={2014},
  volume={25},
  number={10},
  pages={2561-2570},
  doi={10.1109/TPDS.2013.252}}

@ARTICLE{IEEE421516,
  author={},
  journal={IEEE Std 421.5-2016}, 
  title={{IEEE} Recommended Practice for Excitation System Models for Power System Stability Studies}, 
  year={2016},
  volume={},
  number={},
  pages={1-207},
  doi={10.1109/IEEESTD.2016.7553421}}

@misc{Llama24,
  title         = {The {L}lama 3 Herd of Models},
  author        = {Aaron Grattafiori and others},
  year          = {2024},
  eprint        = {2407.21783},
  archiveprefix = {arXiv},
  primaryclass  = {cs.AI},
  url           = {https://arxiv.org/abs/2407.21783}
}

@article{Zhou22,
  author   = {Zhou, Yichen and Wu, Jianwei and Li, Hongyu and Li, Yonggang},
  journal  = {IEEE Trans. Power Syst.},
  title    = {Analysis of Nonlinear Characteristics for Forced Oscillation Affected by Quadratic Nonlinearity},
  year     = {2022},
  volume   = {37},
  number   = {1},
  pages    = {804-807},
  doi      = {10.1109/TPWRS.2021.3120589}
}

@misc{LiLi25,
  title         = {{AI} Load Dynamics--A Power Electronics Perspective},
  author        = {Yuzhuo Li and Yunwei Li},
  year          = {2025},
  eprint        = {2502.01647},
  archiveprefix = {arXiv},
  primaryclass  = {cs.AR},
  url           = {https://arxiv.org/abs/2502.01647}
}

@article{Milano26,
author = {Jiménez-Ruiz, Alberto and Milano, Federico},
title = {Data Centre Model for Transient Stability Analysis of Power Systems},
journal = {IET Gener. Transm. Distrib.},
volume = {20},
number = {1},
pages = {e70381},
doi = {https://doi.org/10.1049/gtd2.70381},
eprint = {https://ietresearch.onlinelibrary.wiley.com/doi/pdf/10.1049/gtd2.70381},
year = {2026}
}

@ARTICLE{Song23,
  author={Song, Yue and Hill, David J. and Liu, Tao and Zhang, Xinran},
  journal={IEEE Trans. Autom. Control}, 
  title={Impasse Surface of Differential–Algebraic Power System Models: An Interpretation Based on Admittance Matrices}, 
  year={2023},
  volume={68},
  number={10},
  pages={6224-6231},
  doi={10.1109/TAC.2022.3230762}}

@ARTICLE{Garcia22,
  author={Garcia, Norberto and Romero, Maria Luisa and Acha, Enrique},
  journal={IEEE Trans. Power Syst.}, 
  title={Jacobian-Free {P}oincaré-{K}rylov Method to Determine the Stability of Periodic Orbits of Electric Power Systems}, 
  year={2022},
  volume={37},
  number={1},
  pages={429-442}}

@ARTICLE{Praprost96,
  author={Praprost, K. L. and Loparo, K. A.},
  journal={IEEE Trans. Autom. Control}, 
  title={A stability theory for constrained dynamic systems with applications to electric power systems}, 
  year={1996},
  volume={41},
  number={11},
  pages={1605-1617},
  doi={10.1109/9.543998}}

@misc{Choukse25,
  title         = {Power Stabilization for {AI} Training Datacenters},
  author        = {Esha Choukse and others},
  year          = {2025},
  eprint        = {2508.14318},
  archiveprefix = {arXiv},
  primaryclass  = {cs.AR},
  url           = {https://arxiv.org/abs/2508.14318}
}

@ARTICLE{Ko25,
  author={Ko, Min-Seung and Zhu, Hao},
  journal={IEEE Trans. Power Syst.}, 
  title={Wide-Area Power System Oscillations from Large-Scale {AI} Workloads}, 
  year={2026},
  volume={},
  number={},
  pages={1-14},
  doi={10.1109/TPWRS.2026.3685506}}

@INPROCEEDINGS{Biswas25,
  author={Biswas, Shuchismita and Varghese, Antos C. and Chatterjee, Kaustav and Nekkalapu, Sameer and Ross, Brett and Follum, Jim},
  booktitle={2026 IEEE/PES Transmission and Distribution Conference and Exposition (T\&D)}, 
  title={Evaluating the Risk to Bulk Power System Reliability from Large Load Induced Oscillations}, 
  year={2026},
  volume={},
  number={},
  pages={1-5},
  doi={10.1109/TD48022.2026.11562229}}

@article{Masle24,
  author   = {Maslennikov, Slava},
  journal  = {IEEE Trans. Power Syst.},
  title    = {Enhancing the Efficiency of Locating the Oscillation Source},
  year     = {2024},
  volume   = {39},
  number   = {6},
  pages    = {7257-7265},
  doi      = {10.1109/TPWRS.2024.3380529}
}

@article{Estevez22,
  author   = {Estevez, Pablo Gill and Marchi, Pablo and Galarza, Cecilia and Elizondo, Marcelo},
  journal  = {IEEE Trans. Power Syst.},
  title    = {Complex Dissipating Energy Flow Method for Forced Oscillation Source Location},
  year     = {2022},
  volume   = {37},
  number   = {5},
  pages    = {4141-4144},
  doi      = {10.1109/TPWRS.2022.3184119}
}

@article{YChen25,
  author   = {Chen, Yunfei and Fan, Ziyu and Gregory, David and Zhou, Xiaoyao and Rabbani, Ronak},
  journal  = {IEEE Access},
  title    = {A Survey of Oscillation Localization Techniques in Power Systems},
  year     = {2025},
  volume   = {13},
  number   = {},
  pages    = {28836-28860},
  doi      = {10.1109/ACCESS.2025.3540318}
}

@article{Rosta94,
  author   = {Rostamkolai, N. and Piwko, R. J. and Matusik, A. S.},
  journal  = {IEEE Trans. Power Syst.},
  title    = {Evaluation of the impact of a large cyclic load on the {LILCO} power system using time simulation and frequency domain techniques},
  year     = {1994},
  volume   = {9},
  number   = {3},
  pages    = {1411-1416},
  doi      = {10.1109/59.336123}
}

@article{Vanness66,
  author   = {Van Ness, J. E.},
  journal  = {IEEE Trans. Power App. Syst.},
  title    = {Response of Large Power Systems to Cyclic Load Variations},
  year     = {1966},
  volume   = {PAS-85},
  number   = {7},
  pages    = {723-727},
  doi      = {10.1109/TPAS.1966.291699}
}

@techreport{wea24,
  author      = {{International Energy Agency (IEA)}},
  title       = {World Energy Outlook 2024},
  institution = {IEA},
  address     = {Paris},
  year        = {2024},
}

@article{RMS_EMT,
  author   = {Lara, Jose Daniel and Henriquez-Auba, Rodrigo and Ramasubramanian, Deepak and Dhople, Sairaj and Callaway, Duncan S. and Sanders, Seth},
  journal  = {IEEE Trans. Power Syst.},
  title    = {Revisiting Power Systems Time-Domain Simulation Methods and Models},
  year     = {2024},
  volume   = {39},
  number   = {2},
  pages    = {2421-2437},
  doi      = {10.1109/TPWRS.2023.3303291}
}

@book{Seydel10,
  author    = {R{\"u}diger Seydel},
  title     = {Practical Bifurcation and Stability Analysis},
  edition   = {3rd},
  series    = {Interdisciplinary Applied Mathematics},
  volume    = {5},
  publisher = {Springer},
  year      = {2010},
  doi       = {10.1007/978-1-4419-1740-9}
}

@book{Kuznetsov04,
  author    = {Yuri A. Kuznetsov},
  title     = {Elements of Applied Bifurcation Theory},
  edition   = {3rd},
  series    = {Applied Mathematical Sciences},
  volume    = {112},
  publisher = {Springer},
  year      = {2004},
  doi       = {10.1007/978-1-4757-3978-7}
}

@article{Hiskens00,
  author  = {I. A. Hiskens and M. A. Pai},
  journal = {IEEE Trans. Circuits Syst. I, Fundam. Theory Appl.},
  title   = {Trajectory Sensitivity Analysis of Hybrid Systems},
  year    = {2000},
  volume  = {47},
  number  = {2},
  pages   = {204-220},
  doi     = {10.1109/81.828574}
}

@article{Venkatasubramanian95,
  author  = {V. Venkatasubramanian and H. Sch{\"a}ttler and J. Zaborszky},
  journal = {IEEE Trans. Autom. Control},
  title   = {Local Bifurcations and Feasibility Regions in Differential-Algebraic Systems},
  year    = {1995},
  volume  = {40},
  number  = {12},
  pages   = {1992-2013},
  doi     = {10.1109/9.478226}
}

@book{Oppenheim1997,
  author    = {Oppenheim, Alan V. and Willsky, Alan S. and Nawab, S. Hamid},
  title     = {Signals and Systems},
  edition   = {2},
  publisher = {Prentice Hall},
  address   = {Upper Saddle River, NJ},
  year      = {1997},
  isbn      = {9780138147570}
}

\end{document}